\documentclass[preprint,12pt]{elsarticle}
\usepackage{amssymb}
\newcounter{bla}

\journal{Computer Physics Communications}

\usepackage{amsmath}
\usepackage{graphicx}
\usepackage[colorinlistoftodos]{todonotes}
\usepackage{bm}
\usepackage{subcaption}
\usepackage{multicol}
\usepackage{verbatim}
\usepackage{cprotect}
\usepackage{enumitem}
\usepackage{appendix}
\usepackage{hyperref}
\usepackage{fancyvrb}
\usepackage{xcolor}

\def\RE{{\rm Re}}
\def\IM{{\rm Im}}

\def\kv{{\bf{k}}}
\def\xu{\hat{\bf{x}}}
\def\yu{\hat{\bf{y}}}
\def\zu{\hat{\bf{z}}}

\def\kpt{  \kv'_{\rm T} }

\def\kptkpt{{\kv'}^2_{\rm T}}

\def\bt {b\T}
\def\bl {b\L}
\def\T{_{\rm T}}
\def\L{_{\rm L}}

\def\be{\begin{equation}}
\def\ee{\end{equation}}

\def\pythiatpz{\texttt{StringSpinner}}
\def\kperp{\textbf{k}_{\perp}}
\def\collff{H_{1q}^{\perp h}}
\def\iff{H_{1q}^{\sphericalangle }}
\def\pythia{PYTHIA}
\def\zgns{\hat{\textbf{Z}}}
\def\ygns{\hat{\textbf{Y}}}
\def\xgns{\hat{\textbf{X}}}

\usepackage{graphicx}

\begin{document}
\begin{frontmatter}
\title{StringSpinner - adding spin to the PYTHIA string fragmentation}
\author[a]{Albi Kerbizi\corref{author}}
\author[b]{Leif L\"onnblad}

\cortext[author] {Corresponding author.\\\textit{E-mail address:} albi.kerbizi@ts.infn.it}
\address[a]{Trieste Section of INFN, University of Trieste, Dept. of Physics, 34127 Trieste, Italy}
\address[b]{Department of Astronomy and Theoretical Physics, Lund University, Sweden}


\begin{abstract}
    \pythiatpz{} is a plugin for the introduction of the spin effects in the hadronization part of the PYTHIA 8 event generator. The spin effects are generated for pseudoscalar meson production by propagating the quark polarization along the fragmentation chain according to the rules of the quantum mechanical string+${}^3P_0$ model. By using parametrizations for the transversity parton distribution functions, \pythiatpz{} allows to simulate the polarized semi inclusive deep inelastic scattering process on protons and neutrons. The instructions to use \pythiatpz{} and to modify the parametrizations of the transversity distribution functions are presented, as well as an example of the calculation of the Collins asymmetries as obtained from simulations of the transversely polarized semi inclusive deep inelastic scattering process.
\end{abstract}
\end{frontmatter}
{\bf PROGRAM SUMMARY} 
\begin{small}
\noindent \\
{\em Program Title: StringSpinner}                                          \\
{\em CPC Library link to program files:} \\
{\em Developer's repository link: https://gitlab.com/albikerbizi/stringspinner.git}  \\
{\em Code Ocean capsule:} \\
{\em Licensing provisions: GNU GPL v2 or later}   \\
{\em Programming language: \rm{C++}, \rm{Fortran}} \\
{\em Nature of problem: To analyse and interpret the experimental data on the study of the nucleon spin structure, as well as to make predictions for future experiments simulations are needed.}\\
{\em Solution method: Introduce spin effects in the hadronization part of a complete Monte Carlo event generator, largerly used in the community of high energy particle physicists.}\\
\end{small}
\section{Introduction}
The nucleon structure can be described in QCD by transverse momentum dependent parton distribution functions (PDFs) which encode information on correlations between the intrinsic transverse momenta of quarks, the spin of the quarks and the spin of the nucleon. In a fast moving nucleon, such functions depend on the lightcone fraction $x$ of the nucleons momentum carried by the quarks, on the intrinsic transverse momentum of the quarks $\kperp$ with respect to the nucleons direction of motion and on the scale $Q^2$ at which the nucleon is probed.

To characterize the collinear ($\kperp$-integrated) structure of the nucleon at leading twist only three PDFs are needed: the number density $f_1^q$ which describes the distribution of unpolarized quarks in an unpolarized nucleon, the helicity $g_1^q$ which describes quarks with longitudinal polarization in a longitudinally polarized nucleon, and the transversity $h_1^q$ which describes transveresely polarized quarks in a transversely polarized nucleon. Among these three collinear PDFs, transversity is the less known and the most difficult to access due to its chiral odd nature.

The quark transverse polarization in nucleons is analysed by using quark polarimeters. Different polarimeters have been proposed. Most of them use the conversion of quarks and gluons in observable hadrons described by means of fragmentation functions (FFs). One such polarimeter is the Collins effect \cite{collins-1993}, namely a left-right asymmetry in the azimuthal distribution of the hadrons produced in the fragmentation process of transversely polarized quarks with respect to the plane defined by the quark momentum and its polarization vector. It is characterized by the chiral odd Collins fragmentation function $\collff$. 
An other mechanism used to analyse the quark transverse polarization is the dihadron effect \cite{Collins:1993kq,Jaffe-Jin-Tang,Bianconi-IFF}, described by the interference fragmentation function (IFF) $\iff$ which is thought to be originated from different physical processes compared to the Collins FF.

A powerful process to access transversity is the semi-inclusive deep inelastic scattering (SIDIS) process $lN\rightarrow l'hX$ where an incoming lepton $l$ scatters off a target nucleon $N$ by exchanging a virtual photon and in the final state the scattered lepton $l'$ and the hadron $h$ are observed in coincidence. In the SIDIS cross section the Collins FF is coupled with the transversity PDF producing a $\sin\phi_C$ azimuthal modulation in the cross section. The azimuthal angle $\phi_C=\phi_h+\phi_S-\pi$ is the Collins angle of the produced hadron, $\phi_h$ being the azimuthal angle of the hadrons momentum transverse with respect to the exchanged virtual photon momentum, and $\phi_S$ the azimuthal angle of the target nucleon transverse polarization $\textbf{S}_{\perp}$ with respect to the exchanged photon momentum. The amplitude of the $\sin\phi_C$ modulation is the Collins asymmetry $A_{Coll}\sim \sum_q e_q^2 h_1^q\otimes H_{1q}^{\perp h}$ given by the convolution between the transversity and the Collins functions over the involved transverse momenta. Extractions of transversity and Collins functions have been performed by combining measurements from Collins asymmetries in SIDIS \cite{COMPASS-collins-sivers,hermes-ssa} with measurements of the corresponding Collins asymmetries in $e^+e^-$ annihilation to hadrons processes \cite{belle-spin-asymmetries}. Examples of such extractions can be found in Refs. \cite{M.B.B,Anselmino:2015,Radici-Bacchetta:2018}.

A useful tool to connect the theoretical and experimental studies on the nucleon structure is a Monte Carlo event generator (MCEG). MCEGs provide a full event description in all the allowed phase space and can be used to guide the phenomenological interpretation of the experimental data or to define new analyses to be performed in the data. One of the advantages of MCEGs is the possibility to access the correlations among particles produced in the same event, difficult to describe theoretically.
At present, however, a complete MCEG implementing the quark spin degree of freedom in the hadronization part is still lacking.

In this article we present \pythiatpz{}, a program module that can be added to PYTHIA 8 \cite{pythia8} as a plugin for the simulation of the polarized SIDIS process, with the aim to making the code available to the comunity. \pythiatpz{} allows to emulate the recursive string+${}^3P_0$ model \cite{kerbizi-2019} of polarized quark fragmentation in the hadronization part of \pythia{}. The string+${}^3P_0$ model is based on the Lund Model \cite{Lund1983} of string hadronization already implemented in PYTHIA, with an additional assumption that the $q\bar{q}$ pairs at string breakups are produced with the vacuum quantum numbers $L=0$, $S=1$ and $J=0$. In addition to the phenomenological parameters of the Lund Model, the $^3P_0$ model introduces only one further complex parameter responsible for the Collins and other spin effects.

The idea behind \pythiatpz{} is to reweight the unpolarized hadronization procedure in \pythia{} to reproduce the string+${}^3P_0$ model, where applicable, to enable the study of the fragmentation of polarized quarks in realistic hadronic final states. Examples of such studies is to measure the Collins effect in transversely polarized quarks, and the effects in the SIDIS process of an unpolarized lepton off a polarized target. The spin effects are presently simulated for the production of pseudoscalar mesons without initial state radiation (ISR) and final state radiation (FSR).
Parametrizations of the transversity distributions for the valence $u$ and $d$ quarks have been implemented and can be used to calculate the polarization vector of the struck quark. They are defined as \verb|C++| functions in a separate file and can be easily modified by the user. This package can therefore be used to simulate the Collins and the di-hadron transverse spin asymmetries (TSAs) or, alternatively, the analysing powers (APs), namely spin asymmetries considering fully polarized quarks.
Some preliminary results obtained with \pythiatpz{} were presented in Ref. \cite{kerbizi-lonnblad}. More details on the string+${}^3P_0$ model and on \pythiatpz{} can be found in Ref. \cite{Kerbizi:2020}.

The article is organized as follows. Section \ref{sec:spin effects} explains how the spin effects are introduced in the hadronization part of \pythia. Section \ref{sec:instructions} gives the practical instructions on how to utilize the plugin and to run a main program. Section \ref{sec:further options} explains how to access different options for the simulations, as for instance how to change the parametrizations of the quark transversity distributions. Section \ref{sec:execution} shows how to run the simulations and sec. \ref{sec:example} gives an example of the simulated Collins asymmetries. Finally we draw our conclusions.

\section{Spin effects in PYTHIA for SIDIS}\label{sec:spin effects}
The plugin has been developed and tested with the version PYTHIA 8.2 of the event generator, but will soon also be available for PYTHIA 8.3.
The spin effects in the hadronization part of \pythia{} are introduced by using the \verb|UserHooks| class which is part of \pythia{} 8.2. This class enables a user to modify behaviours in the event generation at predefined points in the code. In particular one is allowed to accept or reject any hadron being produced in the fragementation according to some logic, which in this case is that of the string+${}^3P_0$ model described in the following. The output is the same as the standard \pythia{} output. No new information is added to the \pythia{} Event Record which reports directly the momenta of the final re-weighted hadrons.

When the SIDIS event generation starts, the Bjorken $x$ variable and the virtuality $Q^2$ of the exchanged virtual photon and the flavor of the initial quark $q_A$ are generated by \pythia{} as well as the hard scattering. Then a string between the scattered quark $q_A$ and the target remnant is set up for hadronization. Since ISR and FSR are switched off, this is a simple string stretched between $q_A$ and the target remnant, with no gluons in between.

\subsection{Polarization vector of the fragmenting quark}
To include the spin effects in the hadronization part of PYTHIA, as preliminary step the polarization vector $\textbf{S}_A$ of the scattered quark is calculated in the gamma nucleon reference system (GNS).
The GNS system is defined by taking $\zgns=\hat{\textbf{q}}$, $\ygns=\textbf{l}\times \textbf{l}'/|\textbf{l}\times \textbf{l}'|$ and $\xgns=\hat{\textbf{l}}_{\perp}$. $\textbf{q}=\textbf{l}-\textbf{l}'$ is the exchanged virtual photon momentum, $\textbf{l}$ and $\textbf{l}'$ being the momenta of the incoming and scattered leptons respectively, and $\hat{\textbf{l}}_{\perp}$ is the transverse momentum of the incoming lepton with respect to $\textbf{q}$. The $\ygns$ axis coincides with the normal to the lepton scattering plane. At leading order in $k_{\perp}/Q$ the GNS frame coincides with the string rest system, namely the reference system where the string is at rest and the string axis defines the $\zu$ axis.

The vector $\textbf{S}_A$ can be either chosen as fixed, i.e. along some given direction, or calculated by using the parametrization of the corresponding transversity distribution $xh_1^{q_A}(x,Q^2)$. 
In the latter case, the transverse polarization vector of the scattered quark with respect to $\zgns$ is
\begin{equation}\label{eq:SA_perp}
\textbf{S}_{A\perp} = \frac{h_1^{q_A}(x,Q^2)}{f_1^{q_A}(x,Q^2)}\times D_{\rm{NN}}(y)\times  \left[\textbf{S}_{\perp} - 2 (\textbf{S}_{\perp}\cdot\hat{\textbf{l}}_{\perp})\hat{\textbf{l}}_{\perp}\right].
\end{equation}
This equation takes into account the fact that quarks in a transversely polarized nucleon are partially polarized and that, after the hard scattering, the transverse polarization vector of the scattered quark is depolarized by the depolarization factor $D_{\rm{NN}}=2(1-y)/[1+(1-y)^2]$, $y$ being the fraction of the virtual photon energy carried by the scattered lepton, and is reflected about the normal to the lepton scattering plane.

To evaluate Eq. (\ref{eq:SA_perp}) the following parameterizations for the valence $u$ and $d$ quark transversity PDFs
\begin{eqnarray}\label{eq:transversity pdf}
xh_1^{u^v}(x) = 3.2\,x^{1.28}\,(1-x)^4, & xh_1^{d^v}(x) = -4.6\,x^{1.44}\,(1-x)^4
\end{eqnarray}
are used. These parametrizations are obtained from a fit to the point by point extraction of transversity PDF performed in Ref. \cite{M.B.B}.
The $Q^2$ dependence on $xh_1^{q_A}$ is neglected. The dependence on the intrinsic transverse momentum of the struck quark is assumed to be the same as the distribution used by PYTHIA for the generation of the primordial transverse momentum. Concerning the unpolarized quark distribution $f_1^{q_A}$, the same parameterizations set by the user in \pythia{} are used. As default we consider however the CTEQ5L set by using the option \texttt{PDF:pSet=2}. This gives values for PDFs close to those used in Ref. \cite{M.B.B}.

The corresponding transversity distributions for the neutron are obtained automatically from the proton ones using isospin symmetry, namely $xh_1^{u^v/n}=xh_1^{d^v/p}$ and $xh_1^{d^v/p}=xh_1^{u^v/n}$. The other quarks are taken unpolarized.

Finally the polarization vector of the scattered quark in the string rest frame is taken $\textbf{S}_A=(\textbf{S}_{A\perp},0)$ 
and it is used to parametrize the $2\times 2$ spin density matrix $\rho(q_A)=(\textbf{1}+\boldsymbol{\sigma}\cdot\textbf{S}_A$ )/2, which provides the initial condition for simulating the polarized splittings.

\subsection{Spin effects in PYTHIA hadronization}
After the spin density matrix of the fragmenting quark is set up, \pythia{} starts the usual fragmentation process by chosing randomly if the first splitting (string break-up) has to be taken from the fragmenting quark side or from the target remnant side. Thus a splitting $q\rightarrow h+q'$ is simulated ($q$ is either $q_A$ or the target remnant) by generating a $q'\bar{q}'$ pair and the hadron $h$ is formed by using the standard \pythia{} procedure.

At this point the plugin inspects the produced hadron which is rejected if the splitting has been taken from the remnant side or if $h$ is not pseudoscalar. The former rejection forces the fragmentation chain to evolve from the $q_A$ side towards the remnant side to emulate the recursive string+${}^3P_0$ model of Ref. \cite{kerbizi-2019}, and this restriction is set as default in the plugin. Alternatively, if $h$ is not a pseudoscalar, it is rejected with a probability $1/2$ in order to not change the composition of hadrons in the event, but only the $\kpt$ distributon of the leading pseudoscalars.

If the hadron passes these selections it is accepted with probability
\begin{equation}\label{eq:prob p}
p(\kpt,\textbf{S}_{q})=\frac{1}{2}\times\left[1-\frac{2\IM(\mu)}{|\mu|^2+\kptkpt}\, \textbf{S}_q\cdot\left(\zu\times\kpt\right)\right],
\end{equation}
being always $0<p<1$. Otherwise the hadron is rejected and a new one is tried by \pythia{} until all selections are passed.
The probability $p$ is inspired to the splitting function of the recursive string+${}^3P_0$ model \cite{kerbizi-2019}, $\kpt$ is the transverse momentum of $q'$ in the string rest frame and $\mu$ the complex mass parameter introduced by the ${}^3P_0$ mechanism to parametrize the spin effects and assumed to be equal for all quark flavors. The imaginary part of $\mu$ is responsible for the transverse spin effects in the fragmentation process.

The transverse momentum $\kptkpt$ is generated by PYTHIA following the standard recipe, namely according to the distribution $f_{pyt}(\kptkpt)= (1-c)\,e^{-\kptkpt/\sigma_0^2}+c\,e^{-\kptkpt/\sigma_1^2}$ \footnote{The PYTHIA parameters corresponding to $\sigma_0^2$, $c$ and $\sigma_1^2$ are \texttt{StringPT:sigma}, \texttt{StringPT:enhancedFraction} and \texttt{StringPT:enhancedWidth} respectively. See also Tab. \ref{tab:pythia parameters}.}. This is at variance with the string+${}^3P_0$ model where $\kptkpt$ is generated according to the distribution $f_{{}^3P_0}(\kptkpt)\propto e^{-\bt \kptkpt}(|\mu|^2+\kptkpt)$ \footnote{The PYTHIA choice $f_{pyt}(\kptkpt)$ is equivalent to chose in the string+${}^3P_0$ model the free input function $f_{\rm T}(\kptkpt)=f_{pyt}^{1/2}(\kptkpt)/(|\mu|^2+\kptkpt)^{1/2}$. In the limit for $c=0$ it gives Eq. (27) in Ref. \cite{kerbizi-2019} with $\alpha=0$.}. However, since in PYTHIA it is $c\ll 1$, the distributions of the produced hadrons transverse momenta are the same as in the string+${}^3P_0$ model for $\bt=\sigma_0^{-1/2}$ as shown in Ref. \cite{kerbizi-lonnblad}.

When the hadron is accepted the spin information of $q$ is transferred to $q'$ by calculating the spin density matrix $\rho(q')$ of $q'$ as in the recursive string+${}^3P_0$ model \cite{kerbizi-2019}
\begin{equation}\label{eq:rho(q')}
    \rho(q')=\frac{\big(\mu+\sigma_z\boldsymbol{\sigma}\cdot\kpt\big)\, \rho(q)\, \big( \mu+\sigma_z\boldsymbol{\sigma}\cdot\kpt\big)^{\dagger}}{\rm{tr}\left[ numerator \right]},
\end{equation}
where $\boldsymbol{\sigma}=(\sigma_x,\sigma_y,\sigma_z)$ is a vector of Pauli spin matrices.

This procedure is applied until the condition for the termination of the fragmentation chain is called by \pythia. This last step is treated by \pythia{} without taking into account spin effects.

\section{The program files and instructions}\label{sec:instructions}
The \pythiatpz{} package can be downloaded from \verb|gitlab| \cprotect\footnote{The repository can be found in \verb|https://gitlab.com/albikerbizi/stringspinner.git|.}.
The package contains the following files
\begin{itemize}[noitemsep,topsep=0pt,parsep=0pt,partopsep=0.0pt]
    \item[-] \verb|StringSpinner.h|
    \item[-] \verb|mc3P0.f90|
    \item[-] \verb|Transversity.h|
    \item[-] \verb|dis.cc|
    \item[-] \verb|Makefile|.
\end{itemize}

\verb|StringSpinner.h| is a header file containing the definitions of classes and functions which constitute the core of the plugin. In particular, this file contains the implementation of the \verb|UserHooks| class which allows to simulate the spin effects. 

\verb|mc3P0.f90| contains a \verb|Fortran| module for the calculation of the quark polarization related quantities, as for instance the ${}^3P_0$ inspired probability $p$.

\verb|Transversity.h| is the header file containing the definitions of the quark transversity distributions and can be changed.

\verb|dis.cc| is the main program providing a basic example of running the plugin for the simulation of the SIDIS process. It is a modified version of \verb|main36.cc| shared among the examples of the PYTHIA 8.2 package. 

Finally \verb|Makefile| contains the instructions for the automatic compilation of the previous files and the link with the \pythia{} library.

\subsection{The main program}
\subsubsection{Activation of spin effects}
The main program \verb|dis.cc| starts with
\begin{verbatim}
#include "Pythia8/Pythia.h"
#include "StringSpinner.h"

using namespace Pythia8;

int main() {

	Pythia pythia;
	Event& event = pythia.event;
	SimpleStringSpinner fhooks(pythia);
	.....
\end{verbatim}
Note that \verb|StringSpinner.h| must be included in the main. This allows to construct the \verb|fhooks| class which is a \verb|SimpleStringSpinner| class, the \verb|UserHooks| base class for the implementation of the spin effects. \verb|fhooks| must be given a \verb|Pythia| object as input.
The new methods introduced to deal with spin effects can be called as \verb|fhooks.method()|. In the following the call of a particular method of \verb|fhooks| will be referred to as \verb|method()|. 

The first option which must be specified is if to simulate TSAs or the corresponding APs. TSAs are simulated if the target polarization vector is specified by using the method \verb|setTargetPol()|. This method requires as input a \verb|Vec4| object which spatial components specify the target polarization vector and in particular only the transverse part is considered in actual simulations. More precisely the line
\begin{verbatim}
    fhooks.setTargetPol( Vec4(0,1,0) );
\end{verbatim}
allows to simulate TSAs for a target which polarization vector is \verb|(0,1,0)|, specified by its components along $\xu$, $\yu$ and $\zu$ axes in the laboratory rest frame.

The simulation of APs, instead, can be performed by using the method \verb|void setQuarkPol| \verb|(int id,Vec4 Squark)| which allows to specify the polarization vector \verb|Squark| for the quark identified by \verb|id|. The integer \verb|id| indicates the quark flavor and for this the usual flavor codes of \pythia{} are used, e.g. \verb|1| for a $d$ quark, \verb|2| for a $u$ quark, \verb|-2| to a $\bar{u}$ antiquark etc. The polarization vectors for the unspecified quarks are assumed to be zero.
For instance the following line
\begin{verbatim}
    fhooks.setQuarkPol(2,Vec4(0,1,0));
    fhooks.setQuarkPol(1,Vec4(0,-1,0))
\end{verbatim}
initialises the polarization vector of $u$ quarks along the $\yu$ axis and that of $d$ quarks along $-\yu$ axis in the laboratory frame.
For APs the scattered quark is not depolarized after the hard scattering, i.e. the depolarization factor $D_{\rm{NN}}$ is not taken into account.

The complex mass parameter introduced by the ${}^3P_0$ mechanism can be set up by using the method \verb|setMu(reMu,imMu)|, where the \verb|double| variables \verb|reMu| and \verb|imMu| indicate the real and the imaginary parts of $\mu$ respectively. This must be done before starting the loop over events. The values of the free parameters and other settings of \pythia{} can be modified by the standard procedure of the generator.

\section{Further options}\label{sec:further options}
\subsection{More general final states}
As already mentioned, the spin effects are presently restricted to pseudoscalar meson production and other hadron types are by default simply rejected.
The production of more general final states can be activated with the method \verb|setHadronMode(mode)|. \verb|mode| is a \verb|Mode| variable set by default to \verb|mode=fhooks.reject|, meaning that all hadron types but pseudoscalar mesons are rejected.
Selecting \verb|mode=fhooks.disable| vector mesons and baryons are accepted with probability $1/2$ as long as spin effects are active, but then spin effects are switched off as soon as a vector meson or baryon is produced. The fragmentation chain then continues as in standard \pythia.

This possibility provides more general final states from the point of view of particle content but spin effects are treated in an approximate way. The simulated TSAs could be diluted more, or less, than how much they would really be if all hadron types were correctly treated with spin effects. Moreover, vector mesons and baryons inherit a partial spin effect from the previous splitting where a pseudoscalar meson was produced, but such effect may not be the correct one expected from a true quantum mechanical treatment of such hadrons. Nevertheless this option could turn out be useful in some cases.

\subsection{Splittings from both sides of the string ends}
The same options are available also for the method \verb|setRemnantMode(mode)|. By setting \verb|mode| to \verb|fhooks.reject|, \pythia{} rejects a hadron if it is produced after a string breakup carried from the remnant side. If \verb|mode| is set to \verb|fhooks.disable| then the hadrons produced from the remnant side are accepted and treated as in ordinary \pythia, i.e. without spin effects. In this case spin effects are applied only if the hadron is produced from the struck quark side. This option is more faithful to the \pythia{} procedure of taking splittings with equal probability from both the struck quark and the target remnant sides.
Among the hadrons produced from the remnant side only pseudoscalar mesons are accepted if \verb|setHadronMode| is set to \verb|reject| whereas all hadron types are accepted if it is set to \verb|disable|.

\subsection{Modification of the transversity PDFs}
The parametrizations of the transversity PDFs for the valence $u^v$ and $d^v$ quarks in a proton target are implemented in \verb|Transversity.h| by the functions \verb|uTransvPDF(x,Q2)| and \verb|dTransvPDF(x,Q2)| respectively. These functions are given in input the values of the variables $x$ and $Q^2$ of the current event, and provide as output the values $xh_1^{u^v/p}$ and $xh_1^{d^v/p}$ obtained with the parametrizations given in Eq. (\ref{eq:transversity pdf}). They can however be modified by the user as simple \texttt{C++} functions.

\subsection{Tuning of the parameters}\label{sec:tuning}
The free parameters of the string+${}^3P_0$ model are summarized in Tab. \ref{tab:pythia parameters}. In the first column are reported the parameters in the \pythia{} language. The correspondence with the parameters of the string+${}^3P_0$ model is shown in the second column (see Ref. \cite{kerbizi-2019}). The values of the free parameters are given in the third and last columns, indicated as "${}^3P_0$ tuning" and "\pythia{} tuning". The first three parameters of the ${}^3P_0$ tuning are the same as in Ref. \cite{kerbizi-2019} whereas the complex mass has been obtained by a chi-square minimization using the Collins asymmetries for pions as measured by COMPASS in SIDIS off protons \cite{COMPASS-collins-sivers} with the corresponding asymmetry simulated with \pythiatpz{} assuming vanishing primordial transverse momentum.

To reproduce the COMPASS Collins asymmetries the complex mass $\mu$ has been changed from $(0.42+i\,0.76)\,\rm{GeV}/c^2$ quoted in Ref. \cite{kerbizi-2019} to $(0.78+i\,0.38)\,\rm{GeV}/c^2$. The same value for $\mu$ gives also a satisfactory comparison with the HERMES data \cite{hermes-ssa}. Also, the same value of the complex mass can be used with the \pythia{} tuning as well, the change in the Collins asymmetries being small.

\begin{table}[h!]
\centering
\begin{tabular}{|c|c|c|c|} 
 \hline
 \pythia{} & string+${}^3P_0$ & ${}^3P_0$ tun. & \pythia{} tun.\\ [0.5ex] 
 \hline\hline
 \verb|StringZ:aLund| & $a$ & $0.9$ & $0.3$\\ 
 \verb|StringZ:bLund| & $\bl/\rm{GeV}/c^2)^{-2}$ & $0.5$ & $0.8$ \\
 \verb|StringPT:sigma| & $\bt^{-1/2}/(\rm{GeV}/c)$ & $0.34$ & $0.304$ \\
 \verb|StringPT:enhancedFraction| &  &  & $0.01$\\
 \verb|StringPT:enhancedWidth| &  &  & $2.0$\\
  \verb|reMu| & $\RE{\mu}/(\rm{GeV}/c^2)$ & $0.78$ & $0.78$ \\
   \verb|imMu| & $\IM{\mu}/(\rm{GeV}/c^2)$ & $0.38$ & $0.38$ \\[1ex] 
 \hline
\end{tabular}
\cprotect\caption{Correspondence between the parameters of the fragmentation process in \pythia{} and in the string+${}^3P_0$ model (for the meaning of the parameters see Ref. \cite{kerbizi-2019}), and the corresponding values for the two tunings.}
\label{tab:pythia parameters}
\end{table}
%

\subsection{Accessing event information}
Methods to read spin dependent variables from the current event, as for instance the value of $xh_1^{q_A}$ and the involved polarization vectors are available from the \texttt{SimpleStringSpinner} class. Such methods must be used inside the event loop.
In particular, the value of the transversity PDF $xh_1^{q_A}$ used in the current event can be accessed using the method \verb|fhooks.xh1q()|.
The method \verb|fhooks.STargetGNS()| provides a \verb|Vec4| variable with the target polarization vector in the GNS frame. The polarization vectors of the quark before the hard scattering and after the hard scattering (fragmenting quark) in the GNS reference frame can be read off with the methods \verb|SQuarkGNS()| and \verb|SFragQuarkGNS()| respectively, which also provide a \verb|Vec4| variable as output.

Information on the kinematic quantities of the current event can be accessed by using the \texttt{DISKinematics(Vec4 lin, Vec4 lout, Vec4 hin)} class, which requires in input the four momenta of the incoming lepton, of the scattered lepton and of the target hadron respectively. This class contains as member variables the DIS invariants \verb|Q2|, \verb|W2|, \verb|xB|, \verb|y|, the matrix \verb|GNS| which brings from the current frame to the GNS frame and the matrix \verb|HCM| which brings from the current frame to the rest frame of the exchanged virtual photon and the target hadron. Both \verb|GNS| and \verb|HCM| are \verb|RotBstMatrix| variables.

Finally, information on spin independent quantities such as the flavor of the struck quark, wether it comes from the valence or the sea and the corresponding value of the unpolarized PDF can be accessed via the standard \pythia{} methods.

\section{Execution}\label{sec:execution}

To execute the \pythiatpz{} main program the \verb|Makefile| has to be set up by specifying the path to the \pythia{} installation directory in the variable \verb|PYTHIADIR|. Then calling \verb|make dis| produces the executable \verb|dis| which is now ready to run with \verb|./dis|.
The \verb|make| command produces in addition the files \verb|routines.mod|, \verb|mc3P0.o| which are created as a consequence of the compilation of the \verb|Fortran| module \verb|mc3P0.f90|. 

\section{Example simulation}\label{sec:example}
\begin{figure}[!h]
\centering
\begin{subfigure}{0.48\textwidth}
\centering
  \includegraphics[width=1.0\linewidth]{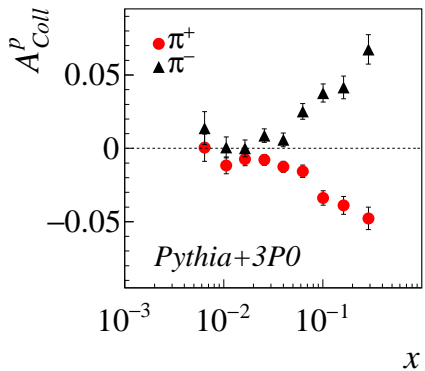}
  \end{subfigure}
\caption{Collins asymmetry as function of $x$ for positive pions (circles) and negative pions (triangles) calculated with Pythia+3P0 in the COMPASS kinematics \cite{COMPASS-collins-sivers}.}
\label{fig: Collins x}
\end{figure}

In Fig. \ref{fig: Collins x} we provide an example of the calculation of the Collins asymmetry for positive pions (circles) and negative pions (triangles) as function of $x$. The asymmetry has been calculated in simulations of $8\,10^6$ SIDIS events in the COMPASS kinematics \cite{COMPASS-collins-sivers}, namely scattering $160\,\rm{GeV}/c$ muons off a fully transversely polarized proton target at rest. We have used the ${}^3P_0$ tuning in Tab. \ref{tab:pythia parameters} and in addition the primordial transverse momentum and the hadron decays have been switched off. \footnote{These settings are set as default in the \texttt{dis.cc} file.}

Concerning the analysis of the simulated data, we applied the same binning and cuts as in the real COMPASS analysis \cite{COMPASS-collins-sivers}. In particular, the cuts $0.1<y<0.9$, $W>5\,\rm{GeV}/c^2$, $0.003<x<0.7$, $z_h>0.2$ and $P_{hT}>0.1\,\rm{GeV}/c$ have been applied, $z_h$ and $P_{hT}$ being the fraction of the exchanged virtual photon energy taken by the observed hadron and the hadron transverse momentum in GNS respectively.

\section*{Conclusions and outlook}
\pythiatpz{} is a plugin to the PYTHIA 8 event generator which allows, for the first time, to simulate the polarized SIDIS process by using the rules of the string+${}^3P_0$ model for the propagation of the quark spin information in the hadronization part of the generator. Parameterizations of transversity PDFs have been introduced for the calculation of the fragmenting quark polarization and can be changed by the user. The spin effects are restricted to the production of pseudoscalar mesons and to string topologies without gluons having switched off ISR and FSR.

\pythiatpz{} can be used for the calculation of the Collins and dihadron transverse spin asymmetries and analysing powers as well. This plugin is useful for the interpretation of the experimental data, to perform multidimensional studies and to make predictions for the future experiments like EIC.

Future developments includes the transition to version 8.3 of PYTHIA, but also to allow for spin dependence in the production of vector mesons, and possibly also baryons. The end-goal is to have a general plug-in to PYTHIA to handle spin-dependent hadronization in any collision system.

\section*{Competing interests}
The authors declare that they have no common competing interests that may have influenced the work described in this article.

\section*{Acknowledgement}
AK thanks Prof. Anna Martin, Prof. Xavier Artru, Prof. Franco Bradamante, Prof. John Collins and Prof. Torbj\"orn Sj\"ostrand for many fruitful and interesting discussions. He acknowledges also interesting discussions with physicists of the Lund group. The authors thank Dr. Markus Diefenthaler, Prof. Stefan Prestel and the LDTMDP group members for fruitful discussions. Part of the work was supported by the
Jefferson Lab LDRD project LDTMDP.

\bibliographystyle{elsarticle-num}
\bibliography{main.bib}

\end{document}